\documentclass[
aps,
prc,
showpacs,
superscriptaddress,
nofootinbib,
floatfix]
{revtex4}
\usepackage{graphicx,
longtable}
\begin{document}
\title{Quasiparticle random phase approximation uncertainties\\ and their correlations in the analysis of $0\nu\beta\beta$ decay}
%
\author{        Amand~Faessler}
\affiliation{   Institute of Theoretical Physics,
				University of Tuebingen, 
               72076 Tuebingen, Germany}
\author{        G.L.~Fogli}
\affiliation{   Dipartimento Interateneo di Fisica ``Michelangelo Merlin,'' 
               Via Amendola 173, 70126 Bari, Italy}
\affiliation{   Istituto Nazionale di Fisica Nucleare, Sezione di Bari, 
               Via Orabona 4, 70126 Bari, Italy}
\author{        E.~Lisi}
\affiliation{   Istituto Nazionale di Fisica Nucleare, Sezione di Bari, 
               Via Orabona 4, 70126 Bari, Italy}
\author{        V.~Rodin}
\affiliation{   Institute of Theoretical Physics,
				University of Tuebingen, 
               72076 Tuebingen, Germany}
\author{        A.M.~Rotunno}
\affiliation{   Dipartimento Interateneo di Fisica ``Michelangelo Merlin,'' 
               Via Amendola 173, 70126 Bari, Italy}
\affiliation{   Istituto Nazionale di Fisica Nucleare, Sezione di Bari, 
               Via Orabona 4, 70126 Bari, Italy}
\author{        F.~\v{S}imkovic}
\affiliation{	Bogoliubov Laboratory of Theoretical Physics, JINR, 
				141980 Dubna, Russia}
\affiliation{	Department of Nuclear Physics,
				Comenius University, 
				Mlynsk\'a dolina F1, SK--842 15 Bratislava, Slovakia}
\affiliation{   Institute of Theoretical Physics,
				University of Tuebingen, 
               72076 Tuebingen, Germany}
\begin{abstract}
The variances and covariances associated to the nuclear matrix elements (NME) of neutrinoless double beta
decay ($0\nu\beta\beta$) are estimated within the quasiparticle random phase approximation 
(QRPA). It is shown that correlated NME uncertainties play an important
role in the comparison of $0\nu\beta\beta$ decay rates for different nuclei, and that they
are degenerate with the uncertainty in the reconstructed Majorana neutrino mass. 
\end{abstract}
\medskip
\maketitle

\section{Introduction \label{SecI}}

The search for the neutrinoless mode of double beta decay ($0\nu\beta\beta$),
\begin{equation}
(Z,\, A)\to (Z+2,\, A) + 2e^- \ ,
\end{equation}
is being vigorously pursued by several experiments using different $(Z,\,A)$ nuclei,
in order to unravel the Dirac or Majorana nature of neutrinos and their absolute mass scale \cite{Av08}.
In a given candidate nucleus $i=(Z,\,A)$, light Majorana neutrinos 
can induce $0\nu\beta\beta$ decay with half-life $T_i$ given by
\begin{equation}
\label{Ti}
T_i^{-1} = G_i\, |M'_i|^2\, m^2_{\beta\beta}
\end{equation}
where $G_i$ is a calculable phase-space factor, $M'_i$ is the $0\nu\beta\beta$ nuclear
matrix element (NME), and $m_{\beta\beta}$ is the 
``effective Majorana neutrino mass,''
\begin{equation}
\label{mbb}
m_{\beta\beta}=\left|\sum_{k=1}^3 m_k\,U^2_{ek}\right|\ ,
\end{equation}
where $m_k$ and $U_{ek}$ are the neutrino masses and the $\nu_e$ mixing matrix elements,
respectively, in standard notation \cite{PDGR}. The NME includes both Fermi (F)
and Gamow-Teller (GT) transitions, plus a small tensor (T) contribution  \cite{Fa98},
\begin{equation}
\label{Mi}
M'_i = \left(\frac{g_A}{1.25}\right)^2
\left(M_i^\mathrm{GT}+M_i^\mathrm{T}-\displaystyle\frac{M_i^\mathrm{F}}{g^2_A}\right)\ .
\end{equation}
In the above expression,
$g_A$ is the effective axial coupling in nuclear matter, not necessarily equal 
to its ``bare'' free-nucleon value $g_A\simeq1.25$. 
With the conventional prefactor $\propto g_A^2$ in Eq.~(\ref{Mi}), the phase space 
$G_i$ becomes $g_A$-independent. In general, all parametric uncertainties (which may be quite
large) are embedded in $|M'_i|$ \cite{Asse}.

It is widely recognized that a convincing case for $0\nu\beta\beta$
decay must involve independent signals in three or more nuclei \cite{Av08,Zu05}.
For instance,  if the theoretical NME uncertainties could be roughly expressed 
in terms of a single nuisance parameter $p$, then one would need
two independent half-life data $T_1$ and $T_2$, and two relations as Eq.~(\ref{Ti}), to fix both
$p$ and $m_{\beta\beta}$ (up to degeneracies). A third datum $T_3$
would overconstrain the system of equations, providing  a cross-check of the results \cite{Gr02,Bi03,Pe04}. 
A negative check might signal possible new $0\nu\beta\beta$ physics 
beyond light Majorana neutrinos (barring experimental or theoretical mistakes).
Any new $0\nu\beta\beta$ mechanism(s) would then involve at least one more unknown, 
and thus it might require one or more data $(T_4,\, T_5, \dots)$ 
for further cross-checks \cite{Si96,De07,Ge07}. Statistical assessments of the various
options demand realistic estimates of experimental and theoretical uncertainties,
and the analysis of possible degeneracies which, as we shall see, may play a relevant role.

Recently, there has been significant progress towards the reduction (and a better evaluation) 
of $0\nu\beta\beta$ theoretical errors.
Within the quasiparticle random phase approximation (QRPA) \cite{Fa98}, these uncertainties
can be largely kept under control by systematically fixing, in each nucleus, the
particle-particle strength parameter $g_{pp}$ via two-neutrino double beta
($2\nu\beta\beta$) decay rates. In this way, the dispersion of NME values
obtained by varying several QRPA ingredients has been significantly reduced (see \cite{Asse} and
references therein).

However, the estimated NME {\em variances\/} do not exhaust the information needed
to compare $0\nu\beta\beta$ limits (or signals) in different nuclei: 
the NME {\em covariances\/} are important as well. Nonzero NME covariances 
have been implicitly recognized in a few works, e.g.,
by studying the dispersion of NME ratios \cite{Pe04}, and by observing that such a dispersion
may be smaller than for individual NME \cite{De07}. To our knowledge, these observations---implying
positive NME correlations---have not yet 
been sharpened from a statistical viewpoint, despite their relevant
consequences for the comparison of $0\nu\beta\beta$ signals. In a nutshell,
the main points can be illustrated as follows. 
If a finite half-life $T_i$ is measured in a nucleus $i$, the half-life
expected in another nucleus $j$ is
\begin{equation}
\label{Tj}
T_j = T_i\, \frac{G_i}{G_j}\frac{|M'_i|^2}{|M'_j|^2}\ ,
\end{equation}
within (large) NME uncertainties. 
From the experimental viewpoint, the ``most favorable case'' would entail 
the shortest decay timescale $T_j$, namely,  the smallest  
$|M'_i|$ and the largest $|M'_j|$. However, if the two NME uncertainties were positively
correlated (e.g., via a common normalization factor), opposite changes 
of $|M'_i|$ and $|M'_j|$ would be unlikely, thus preventing the occurrence
of the ``experimentally favorable'' case. Moreover, a common shift of the NME for  {\em all\/} nuclei
could always be compensated by an inverse shift in $m_{\beta\beta}$ via Eq.~(\ref{Ti}),
leading to a degeneracy between (correlated) theoretical errors and the Majorana neutrino mass.

The purpose of this paper is to explore and discuss these issues in detail.
In Sec~II we set our notation and conventions. In Sec.~III we present our evaluation of the covariance
matrix for the NME in a set of nuclei. In Sec.~IV we apply our formalism to relevant cases in
the $0\nu\beta\beta$ phenomenology. In Sec.~V we summarize our work and discuss future perspectives. 
An Appendix collects additional details about different theoretical evaluations
of $G_i$ and $|M'_i|$.

\section{Notation and conventions}

In the spirit of Refs.~\cite{Fo04,Fo07,Fo08}, we shall use logarithms of the main $0\nu\beta\beta$ quantities in appropriate
units, namely:
\begin{eqnarray}
\label{taui}
\tau_i   &=&   \log_{10}(T_i/\mathrm{y})\ ,\\
\label{gammai}
-\gamma_i &=&   \log_{10} [G_i/(\mathrm{y}^{-1}\mathrm{eV}^{-2})]\ ,\\
\label{etai}
\eta_i   &=&   \log_{10} |M'_i|\ ,\\
\label{mu}
\mu      &=&   \log_{10} (m_{\beta\beta}/\mathrm{eV})\ ,
\end{eqnarray}
so that Eq.~(\ref{Ti}) is linearized as
\begin{equation}
\label{linear}
\tau_i = \gamma_i - 2\eta_i - 2\mu\ . 
\end{equation}
Central values and errors will be denoted as 
\begin{eqnarray}
\label{Etaui}
\tau_i   &=&   \tau_i^0\pm s_i\ ,\\
\label{Eetai}
\eta_i   &=&   \eta_i^0\pm \sigma_i\ ,\\
\label{Emu}
\mu      &=&   \mu^0\pm \delta\ ,
\end{eqnarray}
the $\gamma_i$ having virtually no uncertainties (see however the Appendix for remarks).
Experimental measurements of the $\tau_i$'s are thus
translated into linear constraints on the unobservable quantity $\mu$,
once the nuclear matrix elements $\eta_i$ and their covariances are given. 

Linearization through logarithms is appropriate to deal with relatively
large NME errors.
For instance, a typical ``factor of two'' uncertainty,
$|M'_i|=|M'^0_i|\times(1^{+1.0}_{-0.5})$, entails
at least two drawbacks: 
($i$)  asymmetric errors are difficult to manage
with usual statistical tools (such as least-squares methods);
($ii$)  the unphysical region
$|M'_i|<0$ is hit at twice the lower error. Both drawbacks are avoided by 
expressing the same ``factor of two'' uncertainty as $\eta_i=\eta_i^0\pm 0.30$. 

Concerning the quantities $\tau_i=\log_{10}(T_i/y)$, at present there is only one claim for a positive
$0\nu\beta\beta$ result by Klapdor {\em et al.} \cite{Kl04,Kl06}
as part of the Heidelberg-Moscow Collaboration: $T_i/\mathrm{y}=2.23^{+0.44}_{-0.31}\times 10^{25}$
at $1\sigma$ \cite{Kl06}. We translate this claimed range as
\begin{equation}
\label{Claim}
\tau_i = 25.355 \pm 0.072\ (1\sigma, \ i={}^{76}\mathrm{Ge})\ ,
\end{equation}
where we have slightly displaced the experimental central value so as  
to reproduce the $1\sigma$ extrema, by construction, with symmetric errors \cite{Asym}. For $n>1$, the asymmetric
$n\sigma$ ranges $T_i/\mathrm{y}=2.23^{+n\cdot 0.44}_{-n\cdot 0.31}\times 10^{25}$ correspond to the 
symmetric ranges $\tau_i=25.355 \pm n\cdot 0.072$ within an acceptable accuracy of 10\%, i.e., within about 
$0.2\sigma$  ($0.3\sigma$) at the level of $2\sigma$ ($3\sigma$) ranges. 

The above arguments, as well as the advantages of using linear relations [Eq.~(\ref{linear})]
and  the associated simple statistics (linear propagation of errors, $\chi^2$ methods), lead us
to assume approximately gaussian errors on 
$\log z_i$, rather than on $z_i$ (where $z_i=T_i$, $|M'_i|$), for the purposes of this work.  
In the future, a better knowledge of the probability distributions of the $z_i$'s 
might warrant a different approach, possibly based on
more refined statistical tools applicable to  generic random variables 
(maximum likelihood methods, MonteCarlo simulations). However, our 
main results do not crucially depend on these subtle aspects.

\section{NME UNCERTAINTIES AND THEIR CORRELATIONS}

In this Section we discuss estimates for the nuclear matrix elements $\eta_i$,
in terms of central values $\eta_i^0$,
errors $\sigma_i$ and  correlations $\rho_{ij}$, 
for a set of eight  $0\nu\beta\beta$ candidate nuclei: 
$i={}^{76}$Ge, $^{82}$Se, $^{96}$Zr, $^{100}$Mo, $^{116}$Cd, $^{128}$Te, $^{130}$Te, and $^{136}$Xe. We remind that the associated covariance matrix is $\mathrm{cov}(\eta_i,\,\eta_j) = \rho_{ij}\sigma_i\sigma_j$, whose diagonal elements coincide with the variances $\sigma^2_i$.

\subsection{Numerical evaluation of QRPA uncertainties}

Our estimates are based on a large set of QRPA calculations \cite{Asse,Anat} which include $2\times 2\times 3\times 2=24$ 
variants in the input ingredients, namely:
($i$) two values for the axial 
coupling: $g_A=1.25$ (bare) and $g_A=1.00$ (quenched); ($ii$) two approaches to short-range
correlations (s.r.c.): the so-called Jastrow-type s.r.c., and the unitary correlation operator method (UCOM); 
($iii$) three sizes for the model basis: small, 
intermediate and large; ($iv$) two many-body models: QRPA and
its renormalized version (RQRPA). All the 24 variants are 
supplemented by errors induced by $g_{pp}$ uncertainties (within 
the experimental  $2\nu\beta\beta$ constraints). Concerning NME error estimates,
we adopt the same conservative approach as in \cite{Anat}, and define
the $1\sigma$ range $\eta^0_i\pm\sigma_i$ as the one embracing the minimum and maximum  
calculated value of $\eta_i$ for each nucleus $i$.
These $\pm1\sigma$ errors are more generous than their formal statistical definition (which would
embrace only $\sim 68\%$ QRPA variants, i.e., $\sim 16$ out of 24). Finally, we 
calculate the correlation index $\rho_{ij}$ between  joint ($\eta_{i},\eta_{j}$) values taken
from the same QRPA sample. In all cases, we also include $g_{pp}$-induced variations. 

Our final results for $\eta_i$, $\sigma_i$  and $\rho_{ij}$ 
are reported in Table~I  (together with the values of the
phase space factors $\gamma_i$, for completeness). 
Figure~1 shows the same results in graphical form, for each couple of different nuclei,
in the plane charted by the coordinates $(\eta_i,\eta_j)$. In each panel we show the ``$1\sigma$ 
error ellipse,'' centered at $(\eta^0,\eta^0_j)$ and with correlation $\rho_{ij}$; 
its projection onto a coordinate axis  
coincide with the $\pm1\sigma_i$ range defined previously. Also shown in each panel is the
set of QRPA calculations used, supplemented by the horizontal and vertical error bars induced
by $g_{pp}$ uncertainties (for a total of 24 ``crosses'' in each plane) .

\begin{table}[t]
\caption{For each nucleus $i$, we report the phase space factor $\gamma_i$, the central value
of the nuclear matrix error $\eta_i$, and the error $\sigma_i$, together with the (symmetric) 
error correlation matrix $\rho_{ij}$, according to the QRPA estimates in this work. See the text for
definitions.}
\begin{ruledtabular}
\begin{tabular}{c|c|c|c|llllllll}
&&&& \multicolumn{8}{c} {correlation matrix $\rho_{ij}$}   \\   
$i$ &  $\gamma_i$ & $\eta_i^0$ & $\sigma_i$ & $^{76}$Ge&$^{82}$Se&$^{96}$Zr&$^{100}$Mo&$^{116}$Cd&$^{128}$Te&$^{130}$Te&$^{136}$Xe \\[1mm]
\hline \noalign{\smallskip}
$^{76}$Ge& $25.517$~~ & 0.635~~ & 0.122~~  & ~1      &        &        &        &        &        &        &    \\[1mm]
$^{82}$Se& $24.870$~~ & 0.571~~ & 0.135~~  & ~0.978  & 1      &        &        &        &        &        &    \\[1mm]
$^{96}$Zr& $24.550$~~ & 0.038~~ & 0.247~~  & ~0.518  & 0.506  & 1      &        &        &        &        &    \\[1mm]
$^{100}$Mo& $24.660$~~ & 0.503~~ & 0.162~~  & ~0.973  & 0.957  & 0.491  & 1      &        &        &        &    \\[1mm]
$^{116}$Cd& $24.622$~~ & 0.404~~ & 0.150~~  & ~0.961  & 0.961  & 0.474  & 0.965  & 1      &        &        &    \\[1mm]
$^{128}$Te& $26.073$~~ & 0.534~~ & 0.154~~  & ~0.947  & 0.968  & 0.515  & 0.916  & 0.930  & 1      &        &    \\[1mm]
$^{130}$Te& $24.674$~~ & 0.498~~ & 0.158~~  & ~0.899  & 0.927  & 0.575  & 0.862  & 0.870  & 0.964  & 1      &    \\[1mm]
$^{136}$Xe& $24.644$~~ & 0.254~~ & 0.187~~  & ~0.805  & 0.846  & 0.663  & 0.747  & 0.773  & 0.898  & 0.916  & 1~~\\[1mm]
\end{tabular}
\end{ruledtabular}
\end{table}

In Fig.~1, the strong, positive correlation among theoretical estimates emerges at a glance.  
The QRPA calculations are mostly scattered along a primary direction (the ellipse major axis) 
with positive slope, essentially 
as a result of variations in the s.r.c.\ model (either Jastrow, blue, or UCOM, red) and,
secondarily, to variations in $g_A$. There is also some dispersion in the orthogonal direction
(ellipse minor axis), which is mainly
due to $g_{pp}$ variations. In general, the overall scatter of QRPA is very well captured by
the ellipses, with the possible
exception of those involving $j={}^{96}$Zr, which are somewhat under-sampled at low $\eta_{j}$.
For this nucleus, the $g_{pp}$ parameter turns out to be extremely close to the so-called
QRPA collapse point, the $\eta_j$ estimates becoming less reliable and more erratic as collapse 
is approached---leading to  large and asymmetric error bars.
For other nuclei, $g_{pp}$ is far from the collapse point and the results are more stable
(with smaller and more symmetric $g_{pp}$ errors), as compared to
${}^{96}$Zr. In conclusion, the correlations $\rho_{ij}$ reported in Table~I appear adequate to characterize 
the scatter of QRPA variants, with the only possible exception of ${}^{96}$Zr, whose estimates
must be taken with a grain of salt. 

We remark that the above estimates, performed within the QRPA, 
include only known and controllable sources of uncertainties. Some of them are peculiar of 
QRPA (e.g., $g_{pp}$), while others are common to any nuclear model (e.g., $g_A$ and the s.r.c.). 
It is not excluded that future developments 
in  nuclear theory and data may suggest the inclusion of further parametric uncertainties, 
most notably those related to deformation and to low-lying $\beta^+$ strengths. 

Indeed, a reliable description of the low-lying $\beta^+$ strengths is a challenging task, which calls for
some improvement of the QRPA calculations. In fact, the $2\nu\beta\beta$-decay matrix element used 
to fix the value of $g_{pp}$ is dominated by contributions of low-lying states of the intermediate nucleus.
A recent study \cite{Sim08} has shown that a better agreement for contributions of low-lying states to the $2\nu\beta\beta$-decay matrix element can be achieved by adjusting the single-particle energies so as to reproduce experimental occupation numbers of neutron and proton valence orbits in $^{76}$Ge and $^{76}$Se. 
For a systematic analysis of this kind of effects one needs 
more experimental data (measuring the neutron and proton occupancies
in particle adding and removing transfer reactions~\cite{Schi08}, measuring the
beta strength distributions in charge-exchange reactions~\cite{Fre08}, etc.) and further
theoretical studies, which go beyond the scope of this paper.

Our results must thus be interpreted as an attempt to quantify conservatively the role of known
QRPA uncertainties, which does not exclude that further corrections 
may be required by future developments in this evolving field of research.

\subsection{Comparison with other estimates and discussion}

In Fig.~2 (in the same coordinate planes of Fig.~1) 
we show our error ellipses at 1, 2 and 3 standard deviations ($\Delta\chi^2=1$, 4, and 9,
respectively), and superpose the latest QRPA results from 
Ref.~\cite{Su08} (dots) and the latest shell-model results from Ref.~\cite{She1,She2} (stars, for the available nuclei).
For each nucleus, these independent $\eta_i$ evaluations
fall within our estimated $3\sigma$ range, $\eta_i^0\pm3\sigma_i$. 
Joint estimates of $(\eta_i,\eta_j)$ for couples of nuclei
appear to be roughly aligned along (or parallel to) the major axis of each ellipse, providing an independent
confirmation of positive correlations between the NME. 
The joint estimates also fall within
our $3\sigma$ ellipses in most cases, with a few moderate exceptions 
in some panels of Fig.~2. We refrain, however, from enlarging our errors (or decreasing their
correlations), in order to accommodate these few outliers within each $3\sigma$ error ellipse.
A motivated revision of our estimates
should be based on a detailed comparison of our probability distributions with analogous ones from
independent calculations---rather than with a few sparse points from the published literature. 

Therefore, it would be useful if other theoretical groups in the $0\nu\beta\beta$ field 
could also present ``statistical samples'' of NME calculations, 
as suggested in this work, so as to provide independent estimates of (co)variances for
their NME estimates. In fact, our (co)variances cannot be directly applied to other NME evaluations which,
in general, do not share the same set of error sources. 
In any case, we stress that  our evaluation of QRPA uncertainties is conservative 
enough to cover the most updated, independent NME calculations within $\pm3\sigma$ 
for each individual nucleus---see Table~VI in the Appendix.
Further work is clearly needed to achieve a better convergence among
the central values estimated in different models and, possibly, to reduce
their associated errors.

Some final remarks are in order. As already mentioned, the high correlation in each panel of Fig.~1 is mainly
due to the fact that, if the s.r.c. model or the $g_A$ parameter are varied, all NME tend to either
increase or decrease jointly. However, the assumption that $g_A$ is the same in all nuclei may be
too strong, as the amount of quenching might change in different nuclei. In particular,
we have shown in \cite{Over} that, by using more data besides $2\nu\beta\beta$ as additional constraints,
the fitted values of $g_A$ is not necessarily constant. Independent variations of $g_A$
in different nuclei would generally weaken the correlations in Fig.~1. Similarly, nucleus-dependent deformations
(ignored in this work) might lead to a further spread of errors and to weaker correlations. 
In general, for any two given nuclei, the more 
different their physics (in terms of $g_{pp}$, $g_A$, deformation, etc.), the weaker their correlation (in terms
of nuclear matrix elements). Our estimated correlations might thus be lowered in the future, should
the standard assumptions in QRPA modeling be relaxed in different ways for different nuclei.
Despite all these caveats, our $\rho_{ij}$ matrix represents at least a first, approximate attempt to
quantify existing correlations of theoretical uncertainties. 
Neglecting $\rho_{ij}$ altogether would definitely lead to worse approximations.

\section{APPLICATIONS}

In this Section we apply the previous results to cases of practical interest,
in order of increasing complexity.

\subsection{An application not involving correlations}

As a first application (not involving correlations), we translate 90\% C.L.\ limits on half-lives
into 90\% limits on the Majorana neutrino mass. We remind that a two-sided 90\% C.L.\ range 
corresponds to $\pm1.64\sigma$ ($\Delta\chi^2=2.7$); therefore, the claim in Eq.~(\ref{Claim}) corresponds to
\begin{eqnarray}
\label{Claim90}
\tau_i &=& \tau_i^0 \pm 1.64\,s_i \nonumber \\
      &=& 25.355 \pm 0.118\ (90\%\ \mathrm{C.L.}, \ i={}^{76}\mathrm{Ge})\ .
\end{eqnarray}
and thus to the following 90\% C.L. range for $\mu$ [as given by Eq.~(\ref{linear})]:  
\begin{eqnarray}
\label{Claim90mu}
\mu \pm 1.64 \delta &=& \frac{1}{2}(\gamma_i-\tau_i^0\pm 1.64\,s_i)-(\eta_i^0 \pm 1.64\,\sigma_i)
                      \nonumber\\
                     &=& -0.554 \pm 0.208\ (90\%\ \mathrm{C.L.}, \ i={}^{76}\mathrm{Ge})\ ,
\end{eqnarray}
where the two errors ($s_i/2$ and $\sigma_i$) have been added in quadrature, being uncorrelated. The
corresponding preferred range for the Majorana neutrino mass is:
\begin{equation}
\label{lowbeta}
m_{\beta\beta}/\mathrm{eV} = [0.17,\,0.45]\ (90\%\ \mathrm{C.L.}, \ i={}^{76}\mathrm{Ge})\ .
\end{equation}

\begin{table}[t]
\caption{Best current limits on half-lives at 90\% C.L. ($T_i>T_i^{90}$ and $\tau_i>\tau_i^{90}$) for
different nuclei $i$, from \protect\cite{Bara}.}
\begin{ruledtabular}
\begin{tabular}{ccccc}
$i$ & $T_i^{90}/\mathrm{y}$ & $\tau_i^{90}$ & Experiment & Ref.  \\[1mm]
\hline
$^{76}$Ge& $1.6\times10^{25}$ & 25.204 & IGEX       & \protect\cite{IGEX} \\
$^{82}$Se& $2.1\times10^{23}$ & 23.322 & NEMO-3     & \protect\cite{NEMO} \\
$^{96}$Zr& $8.6\times10^{21}$ & 21.934 & NEMO-3     & \protect\cite{NEMO} \\
$^{100}$Mo& $5.8\times10^{23}$ & 23.763 & NEMO-3     & \protect\cite{NEMO} \\
$^{116}$Cd& $1.7\times10^{23}$ & 23.230 & Solotvina  & \protect\cite{Solo} \\
$^{128}$Te& $7.7\times10^{24}$ & 24.886 & Geochem.   & \protect\cite{Geoc} \\
$^{130}$Te& $3.0\times10^{24}$ & 24.477 & CUORICINO  & \protect\cite{Cuor} \\
$^{136}$Xe& $4.5\times10^{23}$ & 23.653 & DAMA       & \protect\cite{DAMA} \\
\end{tabular}
\end{ruledtabular}
\end{table}

The best one-sided 90\% C.L.\ limits for various nuclei 
have been recently reviewed in \cite{Bara}, in terms of half-lives at 90\% C.L. ($\tau_i>\tau_i^{90}$), as
reported in Table~II. It is worth noticing that, if former data from the Heidelberg-Moscow
experiment were interpreted as a limit on (rather than a signal of) $0\nu2\beta$ decay, the 90\% C.L.\
bound on the $^{76}$Ge half-life would be $1.9\times10^{25}$~y \cite{Kl01}, slightly stronger than the one placed
by IGEX \cite{IGEX} in Table~II.

The information in Table~II can be transformed into 90\% C.L.
limits of the form $\mu<\mu^{90}$ via
the relation
\begin{eqnarray}
\label{90mu}
\mu &<& \frac{1}{2}(\gamma_i-\tau_i^{90})-\eta_i\nonumber\\
   &<& \frac{1}{2}(\gamma_i-\tau_i^{90})-\eta_i^0+1.64\sigma_i=\mu^{90}\ ,
\end{eqnarray}
where we have linearly added two one-sided limits
at 90\%: an experimental  one($-\tau_i^{90}/2$) and a theoretical one ($-\eta_i^{90}+1.64\sigma_i$). In the 
absence of more detailed information about the (unpublished) probability distribution of 
experimental $\tau_i$'s, this is the
most conservative choice.

Figure~3 shows the results of this exercise, in terms of $m_{\beta\beta}/\mathrm{eV}=10^\mu$. 
The shaded band on the left corresponds to the 90\% C.L. range in Eq.~(\ref{Claim90}), while
the bands on the right are obtained by inserting the $\tau_i^{90}$ limits of Table~II into Eq.~(\ref{90mu}),
except for the very weak limit from $^{96}$Zr which is out of scale. No experiment appears to have
probed the 90\% C.L.\ range preferred by the Klapdor {\em et al.} claim, although IGEX and CUORICINO
have almost reached its lower end. 

It is affirmed in \cite{Cuor} that the CUORICINO limit probes part of the Klapdor {\em et al.} range in
$m_{\beta\beta}$, seemingly in contrast with our results. However, the arguments in \cite{Cuor}
involve a comparison of two {\em different\/}
confidence levels, namely, the 90\% C.L.\ limit
from $^{130}$Te versus the 99.73\% C.L.\ range ($\pm 3\sigma$) from $^{76}$Ge. The 
latter range is a factor of $3\sigma/1.64\sigma=1.83$
wider than the appropriate 90\% C.L.\ range 
used in Fig.~3 (left side), and thus leads to more optimistic conclusions. 
In Ref.~\cite{Fo08} the comparison was consistently made at the same
C.L.\ for both nuclei, but it involved an intermediate step where correlations were not taken into account
(see next subsection), leading again to an optimistic impact for the CUORICINO limit. The analysis proposed 
in this work shows that, actually, neither IGEX nor CUORICINO exclude fractions of the range claimed in \cite{Kl06}
at comparable confidence levels, as far as our estimates for
$\eta_i=\eta^0_i\pm\sigma_i$ (and $\rho_{ij}$) hold.

\subsection{Comparison of half-lives in a couple of nuclei}

Here we consider a more direct comparison via observable half-lives in two nuclei, 
bypassing the unobservable Majorana mass $m_{\beta\beta}$. 
We take two different nuclei $i$ and $j$, characterized by nuclear matrix elements 
$\eta_i=\eta^0_i\pm\sigma_i$ and $\eta_j=\eta^0_j\pm\sigma_j$ with correlation $\rho_{ij}$.
A positive $0\nu\beta\beta$ signal in the first nucleus ($\tau_i=\tau_i^0\pm s_i$) translates
into a favored range for the second nucleus ($\tau_j=\tau_j^0\pm s_j$) as follows.

From Eq.~(\ref{linear})
one obtains, by difference,
\begin{equation}
\tau_j-\tau_i=\Delta_{ij} \pm \epsilon_{ij}\ ,
\end{equation}
where 
\begin{equation}
\Delta_{ij}\equiv \tau_j^0-\tau_i^0= (\gamma_j-\gamma_i)+2(\eta^0_i-\eta^0_j)\ ,
\end{equation}
the error $\epsilon_{ij}$ being obtained by summing in quadrature the correlated
uncertainties associated to the difference $2(\eta^0_i-\eta^0_j)$,
\begin{equation}
\label{epsilon}
\epsilon_{ij}^2= 4(\sigma^2_i+\sigma^2_j-2\rho_{ij}\sigma_i\sigma_j)\ .
\end{equation}
Note that, if the correlation term $-2\rho_{ij}\sigma_i\sigma_j$ were neglected, 
$\epsilon_{ij}$ would be overestimated.
The error $s_j$ associated to $\tau_j^0=\tau_i^0+\Delta_{ij}$ is obtained by summing
in quadrature
the uncorrelated errors $s_i$ and $\epsilon_{ij}$,
\begin{equation}
s_i^2= s_j^2+\epsilon_{ij}^2\ .
\end{equation}
As a result, the error $s_j$ has a nonzero correlation $r_{ij}$ with the
error $s_j$, as given by $r_{ij}\,s_i\, s_j=s^2_i$, namely, 
\begin{equation}
r_{ij}=\frac{s_i}{s_j}\ .
\end{equation}

If we apply the above results to $i={}^{76}$Ge and $j={}^{130}$Te, then
the claim by Klapdor {\em et al.} in Eq.~\ref{Claim}, $\tau_i=25.355\pm 0.072$, 
implies that $\tau_j=24.786\pm 0.161$, with error correlation $r_{ij}=0.447$.
Figure~4 shows the corresponding error ellipse at $1.64\sigma$
(90\% C.L.), in the plane charted by the $0\nu\beta\beta$ half-lives of the nuclei
$i={}^{76}$Ge and $j={}^{130}$Te. The ellipse can be thought as the combined result of two
independent constraints, shown as 90\% C.L.\ bands.   
The horizontal band corresponds
to the experimental claim $\tau_i=25.355\pm (1.64\times 0.072)$. The slanted band corresponds to the
theoretical limits placed by our QRPA estimates on the ratio $T_j/T_i$, namely, $\tau_j-\tau_i=
\Delta\pm (1.64\times \epsilon_{ij})$. 
Note that the projection of the ellipse on the $x$-axis provides the range preferred at 90\% C.L.\ 
for the $^{130}$Te half-life: $T_j/\mathrm{y}=[0.33,\,1.12]\times 10^{25}$. Projections for other
nuclei can be similarly derived, as reported in Fig.~5.

Figure~5 shows the two-sided
ranges preferred by the Klapdor {\em et al.} claim at 90\% C.L. (shaded rectangles on the right), as well as the
one-sided 90\% C.L.\ limits from Table~II (bands on the left), for the same nuclei as in Fig.~3. 
The two-sided limits involve the use of NME errors and correlations, except for $^{76}$Ge, which is 
a purely experimental input. Once more, we see that none of the existing limits can exclude a fraction
of the range favored by Klapdor {\em et al.} \cite{Kl06} at a comparable confidence level, although IGEX
and CUORICINO have almost reached it. The more optimistic claim about the CUORICINO impact in \cite{Cuor} was based 
on a larger favored range for the $^{130}$Te half-life, as obtained by ignoring correlations in the
$\epsilon_{ij}$ estimate of Eq.~(\ref{epsilon}).

We emphasize that the contents of Figs.~3 and 5, although similar, 
are not equivalent. The comparison of experimental
sensitivities in Fig.~3 is made in terms of a derived quantity $(m_{\beta\beta})$, while in Fig.~5
it is directly made in terms of observables ($T_i$). One-sided bounds in Fig.~3 are obtained by
linearly adding 90\% C.L.\ theoretical and experimental limits [Eq.~(\ref{90mu})], while in Fig.~5 
only the latter limits are used; conversely, theoretical errors are used with full correlation information
in the allowed (two-sided) bars of Fig.~5. We think that a comparison in terms of observables, as in Fig.~5, provides
a more faithful representation of the current $0\nu\beta\beta$ decay sensitivities.

We conclude this subsection by discussing the 90\% C.L.\ prospective sensitivities 
(in terms of $T_i$) of the
most promising future $0\nu\beta\beta$ projects. Table~III reports such limits, 
according to the recent review in Ref.~\cite{Bara}.
The values in Table~III are largely beyond the two-sided favored ranges in Fig.~5, except perhaps
for $^{136}$Xe, where the expected sensitivity is only a factor $<2$ beyond the Klapdor {\em et al.}
favored range. This gain may be insufficient if one requires
a more demanding check of the claim, at a confidence level significantly higher than 90\%.
It should be added, however, that all the projects in Table~III expect to proceed in a second phase
of operation with larger exposures and lower backgrounds, improving the quoted sensitivities by, possibly,
another order of magnitude \cite{Bara}.

\begin{table}[t]
\caption{Prospective half-life sensitivities at 90\% C.L. ($T_i^{90}$) for 
different nuclei $i$ in promising future projects, as reported in \protect\cite{Bara}.}
\begin{ruledtabular}
\begin{tabular}{ccc}
$i$ & $T_i^{90}/\mathrm{y}$ & Project   \\[1mm]
\hline
$^{76}$Ge& $2.0\times10^{26}$  & GERDA, MAJORANA         \\
$^{82}$Se& $2.0\times10^{26}$  & SuperNEMO      \\
$^{130}$Te& $2.1\times10^{26}$  & CUORE   \\
$^{136}$Xe& $6.4\times10^{25}$  & EXO       
\end{tabular}
\end{ruledtabular}
\end{table}

\subsection{Combination of half-life data from several nuclei, and degeneracy effects}

Let us consider a future, optimistic situation where $0\nu\beta\beta$ decay is established
in $N$ different nuclei, with measured half-lives 
\begin{equation}
\tau_i=\tau_i^0\pm s_i \ (i=1,\dots,N)\ .
\end{equation}
Assuming
that $0\nu\beta\beta$ decays proceed only through light Majorana neutrino exchange, these measurements
will fix one unknown parameter ($\mu$) via a set of $N$ linear equations analogous to Eq.~(\ref{linear}),
\begin{equation}
\tau_i^0\pm s_i=\gamma_i-2(\eta^0_i\pm\sigma_i)-2\mu \ (i=1,\dots,N)\ ,
\end{equation}
where the experimental errors $s_i$ are, in general, uncorrelated (being obtained in independent 
experiments), while the theoretical errors $\sigma_i$ have nontrivial correlations $\rho_{ij}$ (being
obtained within the same QRPA model). 

This overconstrained system can be solved by the least-squares method, i.e., by minimizing the $\chi^2$
function
\begin{equation}
\chi^2(\mu) = \sum_{ij}(\tau^0_i-\gamma_i+2\eta^0_i+2\mu)W_{ij}(\tau^0_j-\gamma_j+2\eta^0_j+2\mu)\ ,
\end{equation}
where the weight matrix $W_{ij}$ is the inverse of the total covariance matrix (including
experimental and theoretical errors),
\begin{equation}
[W]^{-1}_{ij}= \delta_{ij} s_i s_j + 4\rho_{ij}\sigma_i\sigma_j\ .
\end{equation}
The $\chi^2$ function is quadratic in $\mu$,
\begin{equation}
\chi^2(\mu)=a\mu^2+b\mu+c\ ,
\end{equation}
where 
\begin{eqnarray}
a &=& 4\sum_{ij} W_{ij}\ , \\
b &=& 4\sum_{ij} W_{ij}(\tau^0_i-\gamma_i+2\eta^0_i)\ , \\
c &=&  \sum_{ij} (\tau^0_i-\gamma_i+2\eta^0_i)W_{ij}(\tau^0_j-\gamma_j+2\eta^0_j)\ . 
\end{eqnarray}
The minimum value $\chi^2_{\min}$ and the one-sigma shift $\chi^2_{\min}+1$ are reached
for $\mu=\mu_0$ and $\mu=\mu_0\pm\delta$, respectively, where
\begin{eqnarray}
\mu_0 &=& -\frac{b}{2a}\ , \\
\label{delta} \delta &=& \frac{1}{\sqrt{a}}\ ,\\ 
\chi^2_{\min}&=&c-\frac{b^2}{4a}\ . 
\end{eqnarray}
The fit is acceptable if $\chi^2_{\min}/(N-1)\simeq 1$.  Much higher value of $\chi^2_{\min}$
might signal, e.g., new physics beyond the standard mechanism of $0\nu\beta\beta$ decay via 
light Majorana neutrinos (barring experimental and theoretical mistakes). However, the analysis of
nonstandard mechanisms is beyond the scope of this work.

As a practical example for the standard  $0\nu\beta\beta$ case, we consider decay searches
in each of the four nuclei reported in Table~III, in the hypothesis 
that the true value of $m_{\beta\beta}$ is 0.2~eV (i.e., $\mu=-0.70$), close to the lower end of the
range in Eq.~(\ref{lowbeta}). We assume that
the experiments will measure the expected values for the half-lives $T_i$,
\begin{equation}\label{data}
m_{\beta\beta}/\mathrm{eV}=0.2\  \Longrightarrow\  T_i/\mathrm{y} = \left\{
\begin{array}{ll}
4.43 \times 10^{25} & ({}^{76}\mathrm{Ge})\ , \\
1.34 \times 10^{25} & ({}^{82}\mathrm{Se})\ , \\
1.20 \times 10^{25} & ({}^{130}\mathrm{Te})\ , \\
3.43 \times 10^{25} & ({}^{136}\mathrm{Xe})\ , \\
\end{array}
\right.
\end{equation}
with a fractional uncertainty $\delta T_i/T_i=20\%$ (corresponding to $s_i=0.08$). 
By construction, the best fit to any
combination of these mock
data gives back $\mu_0=-0.7$ and $\chi^2_{\min}$=0. The 
relevant output parameter is then the reconstructed $\mu$ uncertainty, $\delta$, from Eq.~(\ref{delta}).

Table~IV shows the $\delta$ values, for all possible combinations of mock data from
the four nuclei (ranging from a single nucleus to all of them). We comment first the results
in the 6th column, which are obtained by (incorrectly) switching off correlations,
i.e., by setting $\rho_{ij}=\delta_{ij}$, as it is often done in the literature. 

\begin{table}[t]
\caption{Combination of any among the four hypothetical half-life  data $T_i$ in Eq.~(\ref{data})
with experimental
uncertainty $\delta T_i/T_i=20\%$. Results are given in terms of the total $1\sigma$ error $\delta$
on the parameter $\mu=\log_{10}(m_{\beta\beta}/\mathrm{eV})$, including theoretical uncertainties without
and with correlations. Bullets indicate the data included in the evaluation (from 1 to 4 data).}
\begin{ruledtabular}
\begin{tabular}{ccccccc}
\# of data & $^{76}$Ge& $^{82}$Se &$^{130}$Te& $^{136}$Xe& $\delta$ (w/o corr.) & $\delta$ (with corr.)  \\[1mm]
\hline
1 & $\bullet$ & & &                               & 0.128 & 0.128  \\
1 & & $\bullet$ & &                               & 0.141 & 0.141  \\
1 & & & $\bullet$ &                               & 0.163 & 0.163  \\
1 & & & & $\bullet$                               & 0.191 & 0.191  \\
\hline
2 & $\bullet$ & $\bullet$ & &                     & 0.095 & 0.128  \\
2 & $\bullet$ & & $\bullet$ &                     & 0.100 & 0.128  \\
2 & $\bullet$ & & & $\bullet$                     & 0.106 & 0.127  \\
2 & & $\bullet$ & $\bullet$ &                     & 0.107 & 0.141  \\
2 & & $\bullet$ & & $\bullet$                     & 0.114 & 0.141  \\
2 & & & $\bullet$ & $\bullet$                     & 0.124 & 0.163  \\
\hline
3 & $\bullet$ & $\bullet$ & $\bullet$ &           & 0.082 & 0.127  \\
3 & $\bullet$ & $\bullet$ & & $\bullet$           & 0.085 & 0.127  \\
3 & $\bullet$ & & $\bullet$ & $\bullet$           & 0.089 & 0.127  \\
3 & & $\bullet$ & $\bullet$ & $\bullet$           & 0.093 & 0.140  \\
\hline
4 & $\bullet$ & $\bullet$ & $\bullet$ & $\bullet$ & 0.075 & 0.127      
\end{tabular}
\end{ruledtabular}
\end{table}

Without correlations, the error $\delta$ is given by the familiar combination of total errors 
from independent data,
\begin{equation}
\label{nocorr}
\frac{1}{\delta^2}=\sum_i \frac{1}{\sigma_i^2+(s_i/2)^2}\simeq \sum_i \frac{1}{\sigma_i^2}\ ,
\end{equation}
where we have used the fact that any of the $\sigma_i$ is a factor of 3--4 greater
than $s_i/2=0.04$. 
Although the error $\delta$ is dominated by theoretical uncertainties, it decreases
by increasing the data sample (see 6th column of Table~IV), 
as a consequence of (incorrectly) assuming no 
correlations. Formally, the combination of all the four data would then provide the estimate
$\mu=-0.7\pm 0.075$, corresponding to $m_{\beta\beta}\simeq 0.2\pm 0.035$.

Unfortunately, including correlations  
spoils this nice result. Table~IV (last column) shows that, with good approximation, 
the uncertainty $\delta$ cannot be much better than the smallest theoretical
uncertainty $\sigma_i$ among the set of nuclei included in the fit. 
Indeed, even with all four nuclei one obtains $\delta=0.127$,
nearly the same as $\delta=0.128$ from the single nucleus $^{76}$Ge (characterized by 
the smallest theoretical error, $\sigma_i=0.122$). Therefore, regardless of how many accurate
experiments are combined, the final accuracy for our test-case Majorana mass will not be better
than $\mu \simeq -0.7 \pm 0.13$, namely, $m_{\beta\beta}\simeq 0.2\pm 0.06$. 

The degeneracy effect induced by correlations
can be easily understood  in the limiting case of equal and 
completely correlated theoretical errors ($\sigma_i\equiv\sigma$ and $\rho_{ij}\equiv \delta_{ij}$).
In this case, the QRPA uncertainties would reduce to a common shift 
$\eta_i\to \eta_i+\delta$ for all nuclei,
where $\delta\in [-\sigma,\,+\sigma]$ within one standard deviation.
[In Fig.~1, the ellipses would collapse 
to ``segments'' with $45^\mathrm{\circ}$ slope in all panels.]
A common shift of all $\eta_i$ is degenerate with a shift
$\mu\to\mu -\delta$ via Eq.~(\ref{linear}),
\begin{equation}
\label{degen}
\tau_i = \gamma_i - 2(\eta_i+\delta) - 2(\mu -\delta)\ ,
\end{equation}
and, thus, the parameter $\mu$ is affected by 
an irreducible uncertainty $\delta = \sigma$. For unequal NME errors $\sigma_i$,
the most accurate one dominates in equations like Eq.~(\ref{degen}) and thus
\begin{equation}
\label{corr}
\delta \simeq \min\{ \sigma_i\}\ ,
\end{equation}
as anticipated.

The difference between $\delta$ estimates without or with correlations, in the combination of
data from $N$ different nuclei, is striking. 
Without correlations, and for comparable
theoretical uncertainties, the error $\delta$ would scale as $\sqrt{N}$  [Eq.~(\ref{nocorr})].
Including correlations, the error $\delta$ becomes dominated by the single, most accurate NME,
irrespective of $N$  [Eq.~(\ref{corr})]. One should thus reduce not only the size,
but also the correlations of theoretical errors, in order to fully exploit the $m_{\beta\beta}$
sensitivity of future, multiple-isotope $0\nu\beta\beta$ searches.

\subsection{Prospective constraints on the absolute neutrino mass and Majorana phase}

The Majorana mass in $0\nu\beta\beta$ decay ($m_{\beta\beta}$) is one of the most
sensitive probes of the absolute neutrino mass scale $m_\nu$, together with 
the effective neutrino mass in beta decay $(m_{\beta\beta})$ 
and the sum of the three neutrino masses in cosmology ($\Sigma$); see \cite{Fo08} for updated bounds. 
It is tempting
to combine prospective data on $(m_{\beta\beta},\, m_\beta,\,\Sigma)$ in the 
optimistic case of a possible signal ``waiting around the corner'', i.e., for masses
close to the current {\em conservative\/} cosmological bound $\Sigma\lesssim 0.6$ eV \cite{WMAP,Fo08,Me08}:
\begin{equation}
m_1\simeq m_2\simeq m_3\equiv m_\nu \simeq 0.2 \ \mathrm{eV}\ .
\end{equation}
For the sake of simplicity, within current neutrino oscillation phenomenology \cite{Fo08}, we  
approximate the mixing matrix values $U^2_{ei}$  as:
\begin{eqnarray}
U^2_{e1} &\simeq& 0.69\ , \\
U^2_{e2} &\simeq& 0.31\,e^{i\phi}\ ,\\
U^2_{e3} &\simeq& 0\ , 
\end{eqnarray}
where $\phi$ is an unknown  Majorana phase. For nearly degenerate masses it is thus \cite{Fo04}
\begin{eqnarray}
m_{\beta} &\simeq& m_\nu\ , \\
\Sigma &\simeq & 3 m_\nu\ ,\\
m_{\beta\beta} & \simeq & m_\nu\,f \ ,\label{omega}
\end{eqnarray}
with 
\begin{equation}
f \simeq |U^2_{e1}+U^2_{e2}|\in [0.38,\,1]\ ,\label{range}
\end{equation}
where the upper (lower) end of the range is obtained for the CP-conserving case $e^{i\phi}=+1$ ($e^{i\phi}=-1$).

Let us test the above scenario with mock data, having the following
central values and fractional $1\sigma$ errors:
\begin{eqnarray}
m_{\beta} &\simeq& 0.2(1\pm 0.5) \ \mathrm{eV}\ , \label{data1}\\
\Sigma &\simeq & 0.6(1\pm 0.3) \ \mathrm{eV}\  ,\label{data2}\\
m_{\beta\beta} & \simeq & 0.2(1\pm 0.3) \ \mathrm{eV} \ . \label{data3}
\end{eqnarray}
In the above equations, the 50\% uncertainty on $m_{\beta}$ 
corresponds to the smallest $1\sigma$ error  estimated 
for the upcoming $\beta$-decay  experiment KATRIN
($\delta m_\beta\simeq 0.1$~eV) \cite{Katr} . A 30\% uncertainty on $\Sigma$
seems appropriate (and even conservative) for a signal in next-generation cosmological data \cite{Wong,Lesg}.
The putative $30\%$ uncertainty on $m_{\beta\beta}$ reflects the discussion in the previous
subsection.

Combining the ``data'' in Eqs.~(\ref{data1}) and (\ref{data2}), one obtains
\begin{equation}
m_\nu \simeq 0.2(1\pm 0.25)\ ,
\end{equation}
which, together with Eq.~(\ref{omega}) and the ``datum'' in  Eq.~(\ref{data3}), imply 
\begin{equation}
f \simeq 1\pm 0.4\ .
\end{equation}
This result, compared with the range in  Eq.~(\ref{range}), would slightly prefer one
CP-conserving case  ($e^{i\phi}=+1$) over the other  ($e^{i\phi}=-1$), at the 
level of $\sim\!\! 1.5\sigma$. 
Therefore, in an optimistic---but not completely
unrealistic---scenario with degenerate neutrino masses, such as the one considered above, a possible determination
of $m_\nu\sim 0.2$~eV 
with $\sim\!\!25\%$ accuracy (via $m_\beta$ plus $\Sigma$) might be accompanied by some indications about
the Majorana phase $\phi$ (via $m_{\beta\beta}$). In this sense, we feel sympathetic towards more encouraging viewpoints
\cite{Pasc,Depp} 
than those expressed by a ``no-go detection'' for $\phi$ \cite{Nogo}, although a real ``measurement'' of $\phi$
remains undoubtedly very challenging, even in the most favorable scenarios.

\section{Summary and Prospects}

Nuclear matrix elements for $0\nu\beta\beta$ decay are affected by relatively large
theoretical uncertainties. Within the QRPA approach, we have shown that, within a given
set of nuclei, the correlations
among NME errors are as important as their size. We have made a first
attempt to quantify the covariance matrix of the NME, and to understand its effects in
the comparison of current and prospective $0\nu\beta\beta$ results for two or more nuclei. The effects
have been clarified through a series of examples, involving an increasing number of
observables. It turns out that correlations may severely limit the accuracy in the 
reconstruction of 
$m_{\beta\beta}$ from any number of $0\nu\beta\beta$ observations in different nuclei,
due to a degeneracy between NME and $m_{\beta\beta}$ uncertainties. In particular,
the fractional error on $m_{\beta\beta}$ is ultimately dominated by a single fractional 
NME uncertainty (the smallest one, among the set of nuclei considered).
Breaking correlations between different nuclei
is thus an important goal, which requires
constraining (and improving) the theoretical model of each nucleus by means of
many independent data 
(not only $2\nu\beta\beta$ data as currently used). In this way,
systematic effects common to all nuclei may be reduced. Another relevant goal is
to compare correlation estimates in future independent calculations
(e.g., QRPA versus shell-model). While  pursuing
such a long-term theoretical and experimental program, 
a covariance analysis like the one proposed in this work
may represent a useful tool, in order to correctly estimate current or prospective sensitivities
to $0\nu\beta\beta$ decay and to Majorana neutrino parameters.

\acknowledgments

This work is supported in part by the EU ILIAS project. 
The work of G.L.F, E.L., and A.M.R.\ is also supported by the Italian Istituto Nazionale di Fisica 
Nucleare (INFN) and Ministero dell'Istruzione, dell'Universit\`a e della Ricerca 
(MIUR) through the ``Astroparticle Physics'' 
research project. A.F., V.R., and F.\v{S}.\ acknowledge support of the Transregio SFB Project TR27 
``Neutrinos and Beyond'' and 436 SLK 17/298 of the Deutsche
Forschungsgemeinschaft.
\newpage

\section*{APPENDIX}

This Appendix clarifies the role of 
different conventions about the $0\nu\beta\beta$ phase space factor $G_i$, the
axial vector coupling $g_A$, the nuclear matrix elements $M'_i$, and the 
 nuclear radius $R_\mathrm{nucl}$,
in comparison with other authors. An agreement on
common conventions would be desirable in the future, to avoid possible
confusion  or ambiguity (see also \cite{Cowe}).

According to usual definitions, the phase space  $G_i$ contains a factor 
 $\left(g_A^2/R_\mathrm{nucl}\right)^2$, where $R_\mathrm{nucl}=r_0A^{1/3}$. 
In this work, the adopted values of $G_i$ refer to $r_0=1.1$~fm and $g_A=1.25$ \cite{Pant}, while changes of $g_A$ are
conventionally embedded in $M'_i$ (rather than in $G_i$) via the prefactor $(g_A/1.25)^2$ 
in Eq.~(\ref{Mi}) \cite{Asse,Anat}. In order to match such convention,
alternative calculations of
$G_i$  using $r_0=1.2$~fm \cite{Civi,Taka,Voge} must
be rescaled by a factor $f^2_0\simeq 1.2$ (where $f_0\simeq1.1\simeq 1.2/1.1$) \cite{Cowe}.

Table~V compares three different phase-space calculations (in terms of 
$G_i\, m^2_e$), all normalized to the same reference values $g_A=1.25$ and $r_0=1.1$~fm.  One
can notice residual differences of $\sim 5\%$ between the results of \cite{Pant} and \cite{Taka,Voge},
and of $\sim 10\%$ between those of \cite{Pant} and \cite{Civi}, presumably due to different approximations
used to evaluate the electron wave function and the screening corrections. 
In our opinion, a typical uncertainty for the computed $G_i$ values may be
estimated as $\pm 5\%$, corresponding to a variation 
 $\delta\gamma_i\simeq \pm 0.02$ for the $\gamma_i$ values in  Table~I. Such minor error, being
much smaller than the theoretical and experimental uncertainties  considered in this
work ($\delta \gamma_i\ll\sigma_i,\,s_i$), has been ignored---but it might become more important in the future.

Concerning the 
nuclear matrix elements $|M'_i|$, the values calculated in \cite{Su08} (QRPA) and \cite{She1,She2} 
(shell model) refer to $r_0=1.2$~fm,
and must be rescaled by a factor $1/f_0$ for comparison with the NME used in this work.
Furthermore, since the values in \cite{Su08} do not embed the prefactor $(g_A/1.25)^2$, they must be 
rescaled by another factor $1.25^{-2}$ in the subcase $g_A=1$;
this further rescaling is not necessary for the NME values in \cite{She2}.
Table~VI reports the rescaled values of $|M'_i|$ from \cite{Su08} and \cite{She2} (in terms of logarithms $\eta_i$), 
as also used in Fig.~2. These $\eta_i$ values are all
contained within our estimated three-standard-deviation ranges $\eta_i^0\pm 3\sigma_i$, which
are reported in the last two rows of Table~VI.

\begin{table}[t]
\caption{Comparison of $G_i m^2_e$ estimates
(in units of $10^{15}$~y$^{-1}$) for $g_A=1.25$. The second column refers to the 
calculations reported in \protect\cite{Pant} for $r_0=1.1$~fm, as used in this work. The  third and fourth
columns refer to independent estimates \protect\cite{Civi,Taka,Voge} 
for $r_0=1.2$~fm, rescaled by a compensating factor 
$f^2_0=1.2$.}
\begin{ruledtabular}
\begin{tabular}{cccc}
Nucleus  & Ref.~\protect\cite{Pant}   & Refs.~\protect\cite{Taka,Voge}   &  Ref.~\protect\cite{Civi} \\[1mm]
\hline
$^{76}$Ge& 7.93 & 7.67 & 7.57 \\ 
$^{82}$Se& 35.2 & 33.8 & 32.8 \\
$^{96}$Zr& 73.6 & 70.2 & 68.4 \\
$^{100}$Mo& 57.3 & 54.8 & 52.8 \\
$^{116}$Cd& 62.3 & 59.3 & 56.2 \\
$^{128}$Te& 2.21 & 2.20 & 1.99 \\
$^{130}$Te& 55.4 & 53.2 & 49.7 \\
$^{136}$Xe& 59.1 & 56.8 & 52.4 \\
\end{tabular}
\end{ruledtabular}
\end{table}

\begin{table}[t]
\caption{Estimates of $\eta_i=\log_{10}|M'_i|$ for each nucleus, as derived from 
the recent QRPA calculations in \protect\cite{Su08} (see Tab.~1 therein)  and shell-model
calculations in \protect\cite{She2} 
(see Tab.~7 therein) after appropriate rescaling, in order to match the conventions used in this work.
The estimates of  \protect\cite{She2} 
refer only to a subset of nuclei and to $g_A=1.25$.
The s.r.c.\ used (Jastrow or UCOM) are explicitly reported. The last two rows report the upper and lower
ends of our three-standard-deviation ranges $\eta_i^0\pm3\sigma_i$ (for any s.r.c.\ and $g_A$), 
which embrace all the above $\eta_i$
estimates.}
\begin{ruledtabular}
\begin{tabular}{cccrrrrrrrr}
Ref. & s.r.c. & $g_A$ & $^{76}$Ge & $^{82}$Se & $^{96}$Zr & $^{100}$Mo & $^{116}$Cd & 
$^{128}$Te & $^{130}$Te & $^{136}$Xe \\
\hline
\protect\cite{Su08} & Jastrow & 1.00 & 0.471 &  0.313 &  0.261 &  0.312 &  0.331 &  0.396 &  0.374 &  0.222 \\
\protect\cite{Su08} &  UCOM   & 1.00 & 0.582 &  0.427 &  0.400 &  0.451 &  0.435 &  0.531 &  0.501 &  0.335 \\
\protect\cite{Su08} & Jastrow & 1.25 & 0.564 &  0.401 &  0.274 &  0.396 &  0.441 &  0.488 &  0.435 &  0.271 \\
\protect\cite{Su08} &  UCOM   & 1.25 & 0.687 &  0.529 &  0.452 &  0.553 &  0.554 &  0.639 &  0.584 &  0.406 \\
\protect\cite{She2} & Jastrow & 1.25 & 0.320 &  0.297 &        &        &        &  0.328 &  0.285 &  0.204 \\
\protect\cite{She2} &  UCOM   & 1.25 & 0.407 &  0.380 &        &        &        &  0.418 &  0.382 &  0.299 \\
\hline
This work & \multicolumn{2}{l}{Lower limit at $3\sigma$ level}
                                    & 0.269 &  0.166 &$-0.703$&  0.017 &$-0.046$&  0.072 &  0.024 &$-0.307$  \\
This work & \multicolumn{2}{l}{Upper limit at $3\sigma$ level}
                                    & 1.001 &  0.976 &  0.779 &  0.989 & 0.854  &  0.996 &  0.972 &  0.815
\end{tabular}
\end{ruledtabular}
\end{table}

\newpage 


\newpage

\newpage
\begin{figure}[t]
\vspace*{+1.0cm}
\hspace*{0.cm}
\includegraphics[scale=.97]{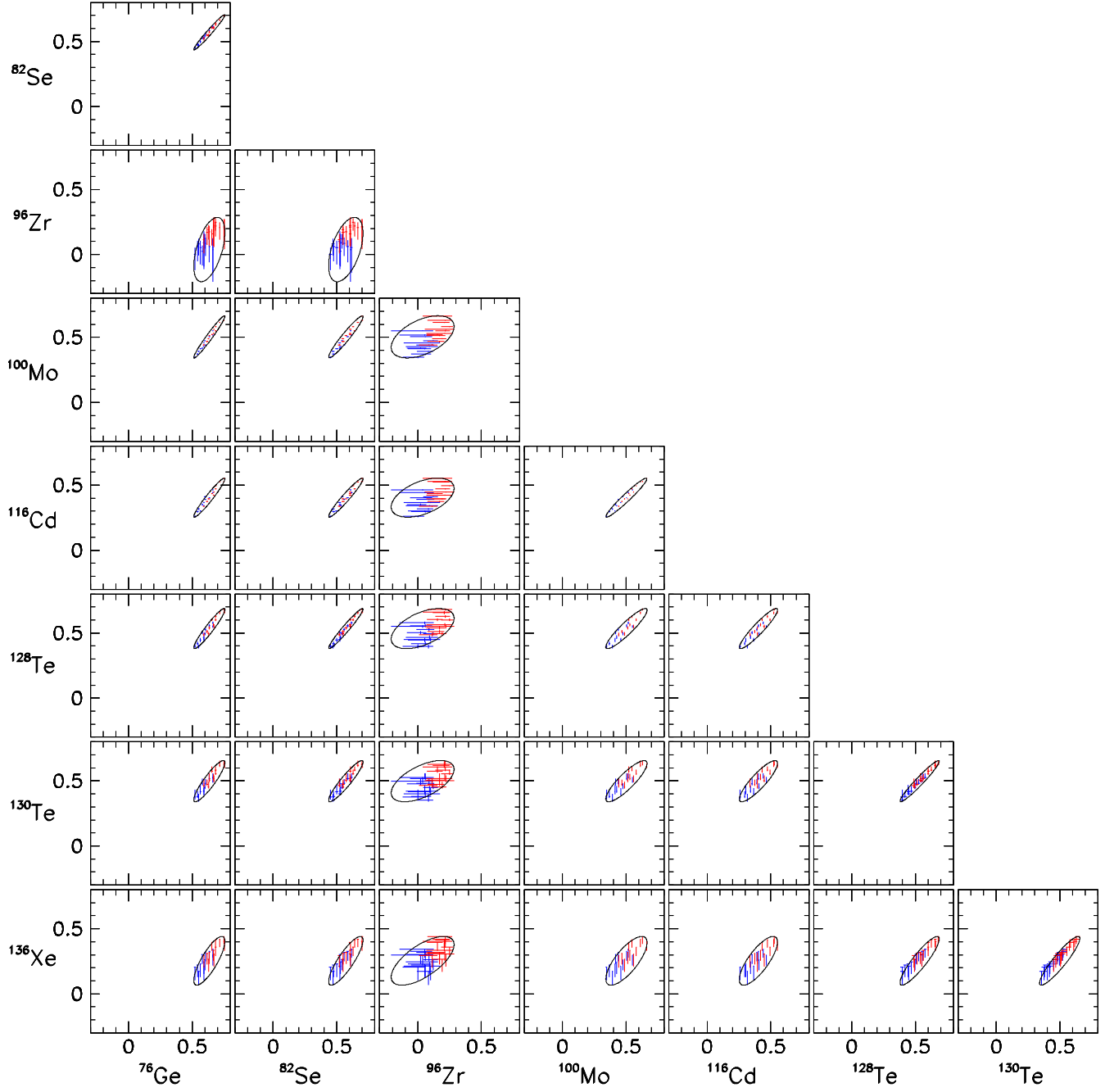}
\vspace*{+1.0cm}
\caption{ \label{f01} Scatter plot of estimated QRPA values for the (logarithms of) 
nuclear matrix elements $(\eta_i,\,\eta_j)$ for each couple of nuclei $(i,\,j)$, together with
the error bars induced by $g_{pp}$ uncertainties. In each panel, also shown is the 
$1\sigma$ error ellipse, conservatively estimated on the basis of the scatter plots. 
See the text for details. Color code for s.r.c: blue (Jastrow), red (UCOM).}
\end{figure}
\newpage
\begin{figure}[t]
\vspace*{+1.0cm}
\hspace*{0cm}
\includegraphics[scale=0.97]{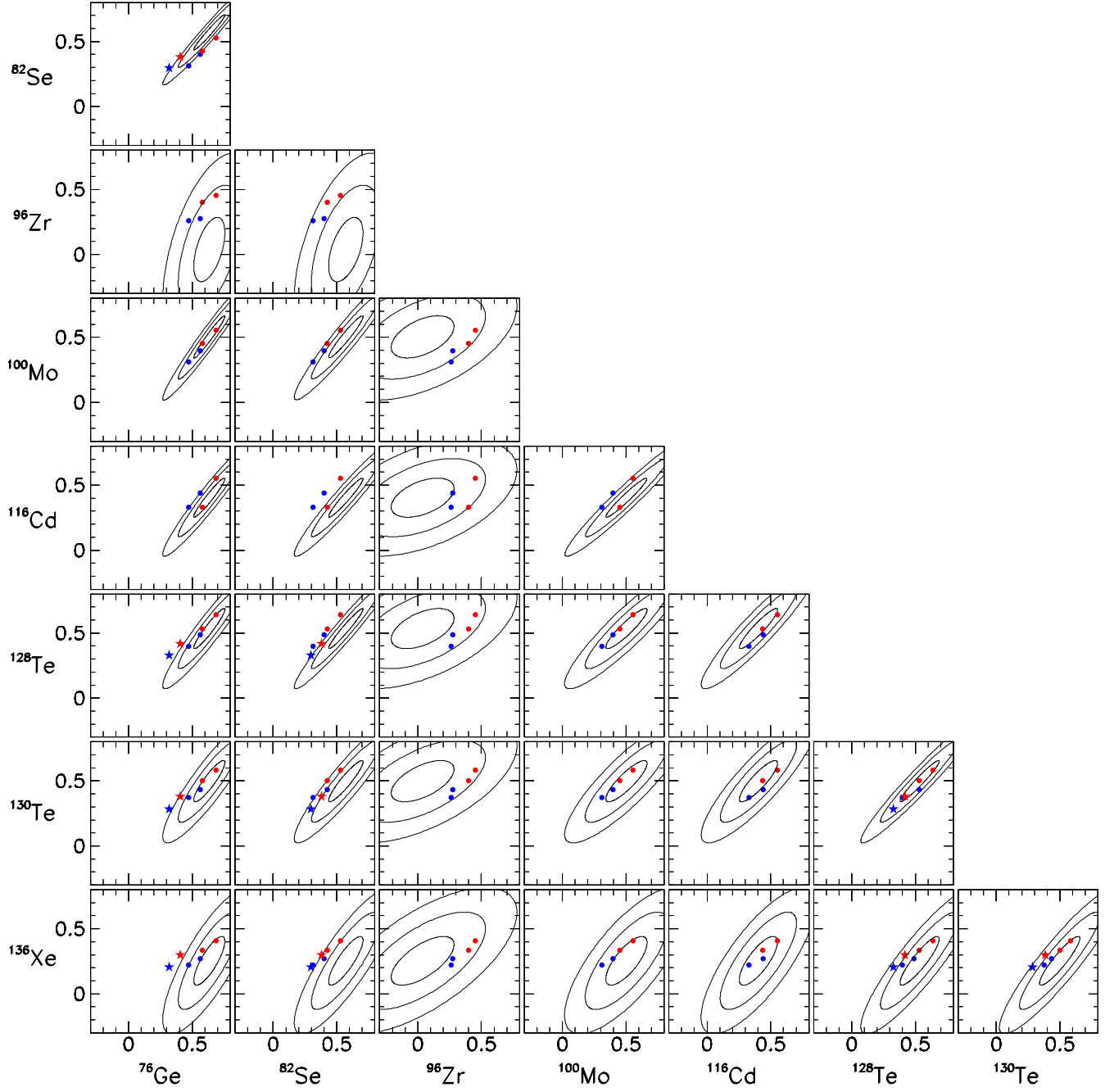}
\vspace*{+1.0cm}
\caption{ \label{f02} Error ellipses at 1$\sigma$, 2$\sigma$ and 3$\sigma$, as derived
from Fig.~1 and compared with independent nuclear matrix element calculations
from \protect\cite{Su08} (QRPA, dots) and 
\protect\cite{She1} (shell model, stars).  Color code for s.r.c: blue (Jastrow), red (UCOM).}
\end{figure}
\newpage
\begin{figure}[t]
\vspace*{+1.0cm}
\hspace*{-0.2cm}
\includegraphics[scale=0.97]{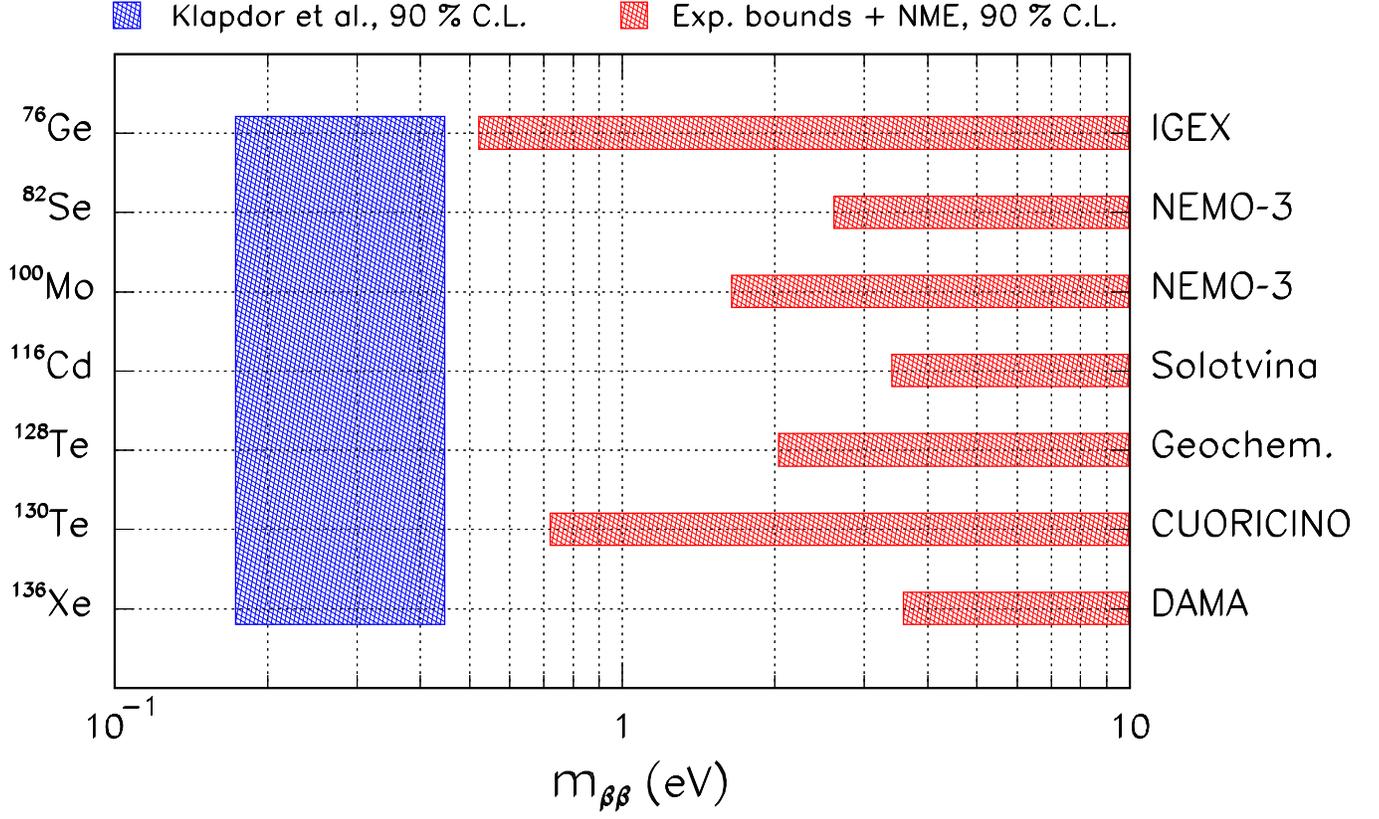}
\vspace*{+1.0cm}
\caption{ \label{f03} Range of $m_{\beta\beta}$ allowed at 90\% C.L. by the $0\nu\beta\beta$ 
claim of \protect\cite{Kl06}, compared with the 90\% limits placed by other experiments. The 
comparison involves the NME and their errors, as estimated in this work.}
\end{figure}
\newpage
\begin{figure}[t]
\vspace*{+1.0cm}
\hspace*{0cm}
\includegraphics[scale=0.97]{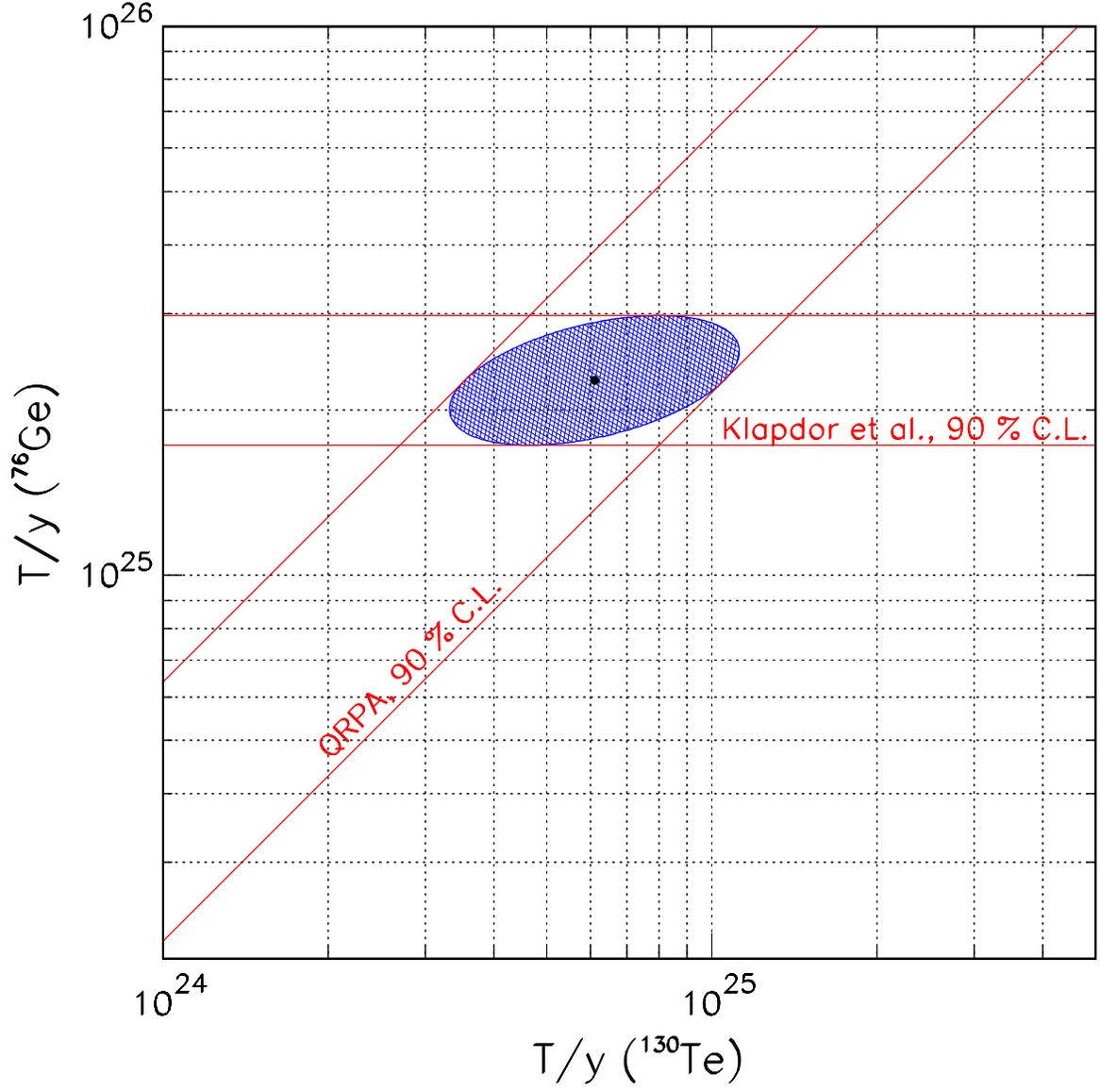}
\vspace*{+1.0cm}
\caption{ \label{f04} 
Example of theoretical and experimental constraints at 90\% C.L., in the plane charted
by the $0\nu\beta\beta$ half-lives of $^{76}$Ge and $^{130}$Te.  Horizontal
band: range preferred by the  $0\nu\beta\beta$ claim of \protect\cite{Kl06}. Slanted band: constraint
placed by our QRPA estimates. The combination of the two constraints 
provides the shaded ellipse, whose projection on the abscissa gives the range preferred at
90\% C.L. for the $^{130}$Te half life.}
\end{figure}
\newpage
\begin{figure}[t]
\vspace*{+1.0cm}
\hspace*{-0.2cm}
\includegraphics[scale=0.97]{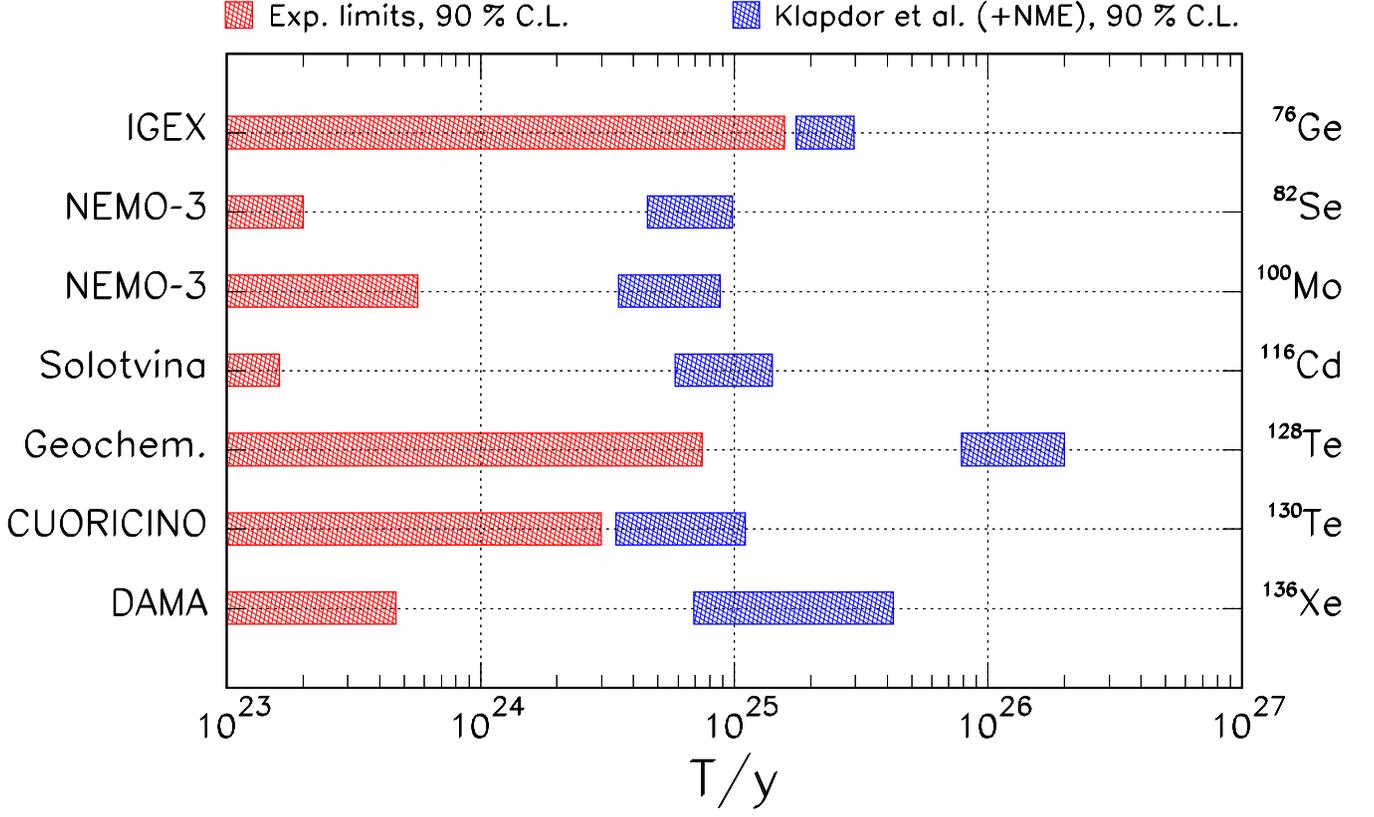}
\vspace*{+1.0cm}
\caption{ \label{f05}
Range of half-lives $T_i$ preferred at 90\% C.L. by the $0\nu\beta\beta$ claim
of \protect\cite{Kl06}, compared with the 90\% limits placed by other experiments. The 
comparison involves the NME and their errors, as well as their correlations, estimated in this work.
}
\end{figure}
\newpage

\end{document}